\documentclass[a4paper,11pt]{article}
\pdfoutput=1 

\usepackage{jinstpub} 
\usepackage[per-mode=symbol,separate-uncertainty=true]{siunitx}
\usepackage{graphicx}
\usepackage{float}
\usepackage{subfig}
\usepackage{lineno}

\title{Position Reconstruction Using Photon Timing for the DEAP-3600
  Dark Matter Experiment}

\author{Y. Chen}
\affiliation{University of Alberta,\\Edmonton, Alberta, T6G 2R3, Canada}

\emailAdd{yu.chen@ualberta.ca}

\abstract{DEAP-3600 is a single-phase liquid argon dark matter
  detector being operated 2 km underground at SNOLAB, Sudbury,
  Canada. The detector consists of 3.3 tonnes of ultra-pure liquid
  argon in a spherical acrylic cryostat instrumented with 255
  photomultiplier tubes. Natural radioactive contamination in the
  acrylic vessel or TPB wavelength shifter can 
  alpha-decay. Reconstruction of the position of the interactions taking
  place in the detector uses information about the number of
  photoelectrons detected in each PMT and when they were
  detected. Including this information in our suite of cuts allows us
  to identify and remove almost all surface background events. A
  method of event position reconstruction emphasizing photon timing is
  presented here.}


\collaboration[c]{on behalf of DEAP collaboration}

\proceeding{LIDINE 2019: Light Detection In Noble Elements\\
  28-30 August 2019\\
  University of Manchester, UK}

\begin{document}
\maketitle
\flushbottom

\section{DEAP-3600 experiment}
\label{sec:intro}
DEAP-3600 is a single-phase liquid argon (LAr) direct-detection dark
matter experiment, located \SI{2}{\km} underground at SNOLAB, Sudbury,
Canada. It has been collecting data since August, 2016. The detector
contains a 3.3 tonne LAr target mass in a spherical acrylic vessel (AV) with an
inner radius of \SI{850}{\mm}, viewed by 255 inward-facing
photomultiplier tubes (PMTs) optically coupled to \SI{45}{\cm}
long acrylic light guides (LGs). The top \SI{30}{\cm} of the AV is
filled with gaseous argon (GAr), and leads to an acrylic neck with
a cooling coil and a set of acrylic flowguides (FGs). The inner
surface of the AV is coated with a \SI{3}{\micro\m} layer of
1,1,4,4-tetraphenyl-1,3-butadiene (TPB) that converts \SI{128}{\nm}
LAr scintillation light to visible wavelengths with a spectrum that
peaks at \SI{420}{\nm}. At these wavelengths, the light can travel
through the AV and LGs and be detected by the PMTs. For a more detailed
description of the detector see~\cite{deap2nd}. 

Fiducialization and surface background rejection rely on position
reconstruction. DEAP-3600 uses two complementary position
reconstruction algorithms. One is a photoelectron (PE)-based
algorithm, which fits the spatial distribution of PMT charge with the
model distribution of PEs detected by each PMT for different test
event positions, pre-trained by a Monte Carlo simulation of the
detector. The other is a time-of-flight-based algorithm, which uses
both spatial and timing information of the hits in the first
\SI{40}{\ns} to fit the position. This should be compared with the
PE-based algorithm that uses both prompt and late light and is based
on the way that intensity drops off with distance. In the WIMP
search analysis with the 231-day exposure dataset published in
2019~\cite{deap2nd}, we have used the PE-based algorithm
for fiducialization, and used the time-of-flight-based algorithm to
remove backgrounds induced by $\alpha$-decays from the AV neck.  

\section{Time-of-flight-based position reconstruction algorithm}
\label{sec:tf2}
\begin{figure}[!tbp]
\centering
\subfloat{}\includegraphics[width=0.45\textwidth]{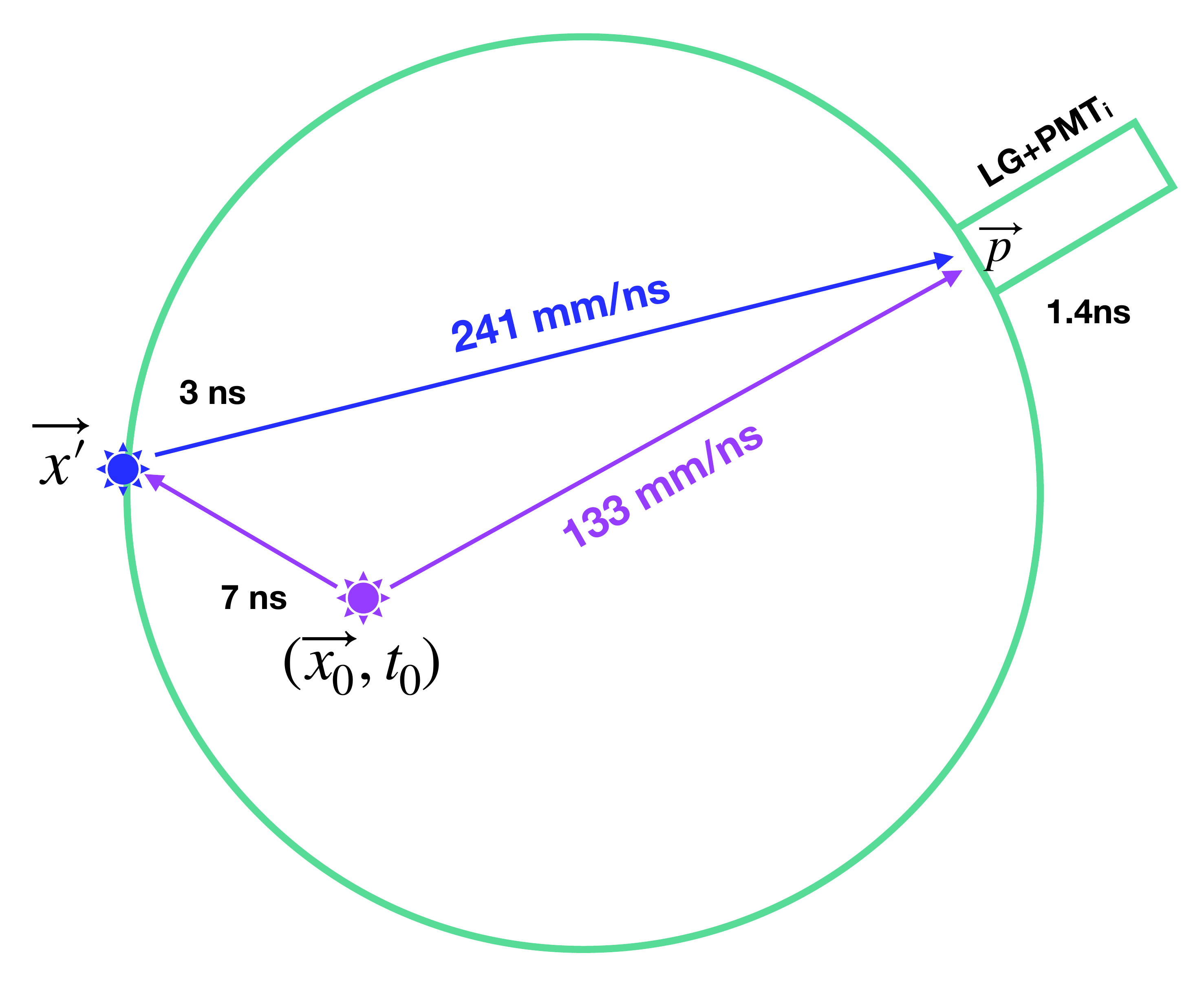}
\subfloat{}\includegraphics[width=0.45\textwidth]{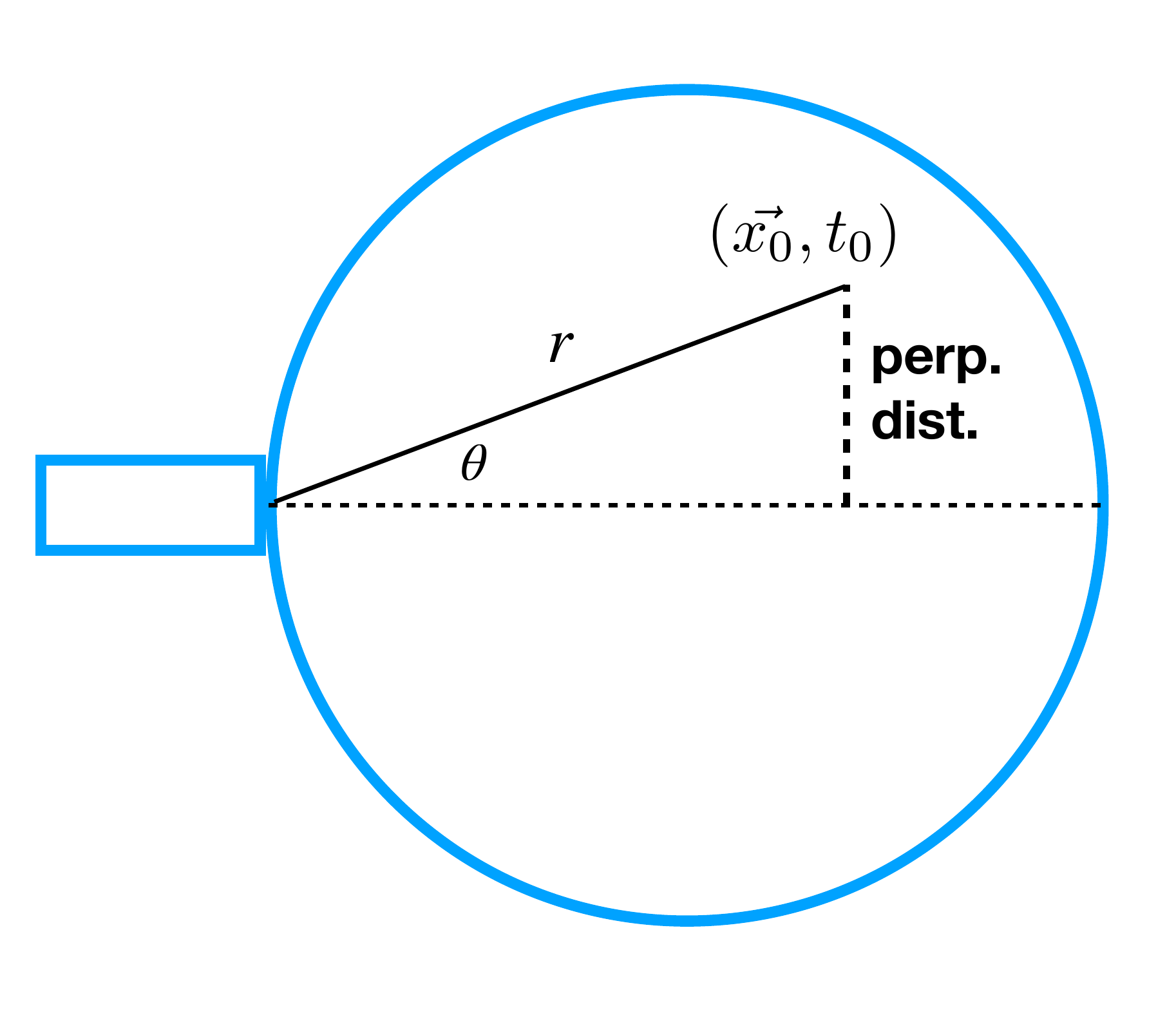}
\caption{\textit{Left:} The light propagation model in the TOF-based
  position reconstruction algorithm. \textit{Right:} The
  coordinate system used to construct the PDF for the TOF-based algorithm. }
\label{fig:scint}
\end{figure}

The time-of-flight (TOF)-based position reconstruction algorithm takes advantage of
the finite speed of light propagation to the PMTs. The group
velocity of UV light at a wavelength of $\SI{128}{\nano\meter}$ in
LAr is known to be $\SI{133\pm1}{\milli\meter\per\nano\second}$
~\cite{DuneOptics}, while that of visible light at a wavelength of
$\SI{420}{\nano\meter}$ is
\SI{241}{\milli\meter\per\nano\second}~\cite{GraceNikkel}. Both values
are used in the TOF-based algorithm. A test with
the values of group velocities varying by 5\% has resulted in
negligible effect on position reconstruction resolution. 

The algorithm is based on an simplified optical model, demonstrated in
the left panel of Figure~\ref{fig:scint}. For a scintillation interaction
event taking place at event time $t_0$ and position $\vec{x_0}$ in
LAr, the arrival time of a photon at a PMT can be calculated as 
\begin{equation}
  t_i = t_0 + \mathrm{TOF}(\vec{p},\vec{x_0})  + \tau,
\end{equation}
in which $\mathrm{TOF}(\vec{p},\vec{x_0})$ is the TOF
from the event vertex $\vec{x_0}$ to position
$\vec{p}$, defined as the center of the LG front surface. $\tau$
spreads by a random time, which is based on three physical
processes: singlet decay with time constant
\SI{7}{\nano\second}~\cite{Hitachi83}, 
TPB re-emission with time constant \SI{3}{\nano\second}, 
and propagation in LG and PMT response with total uncertainty in time
\SI{1.4}{\ns}. The average TOF for light propagation in LG
and the PMT response time are removed by calibration, so we
effectively treat the front surface of LGs as where the light is
detected. The spread due to the LG light propagation and the PMT response
times are included into the random time delay term. The digitizer
samples at \SI{4}{\ns}, but the analog electronics were designed to
allow significantly better resolution, which is dominated by the PMT
response time and has been measured to be \SI{1.4}{\ns}.~\cite{deapdetector}

A primary photon can be absorbed by TPB when it reaches a
point $\vec{x'}$ on the AV surface, re-emitted with the
wavelength shifted to
$\sim\SI{420}{\nano\meter}$, and travel with the greater group velocity
to reach a LG front point $\vec{p}$. The TOF can be calculated as
\begin{equation}
  \label{eq:tof}
  \mathrm{TOF}(\vec{p},\vec{x_0}) =
  \frac{\|\vec{x'} - \vec{x_0}\|}{v_{\mathrm{uv}}} +
  \frac{\|\vec{p} - \vec{x'}\|}{v_{\mathrm{vis}}}, 
\end{equation}
where $v_{\mathrm{uv}}$ and $v_{\mathrm{vis}}$ are the group velocity
of the \SI{128}{\nano\meter} UV light and that of the \SI{420}{\nano\meter}
visible light in LAr, respectively. When a photon of UV light is
reaching the LG front $\vec{p}$ directly from the
scintillation vertex $\vec{x_0}$, it can be considered as
$\vec{x'}=\vec{p}$, then only the TOF of UV contributes to
the above equation. Reflections and scattering of visible photons in
TPB, as well as Rayleigh scattering in the LAr, are neglected. 

\begin{figure}[!tbp]
\centering
\subfloat{}\includegraphics[width=0.49\textwidth]{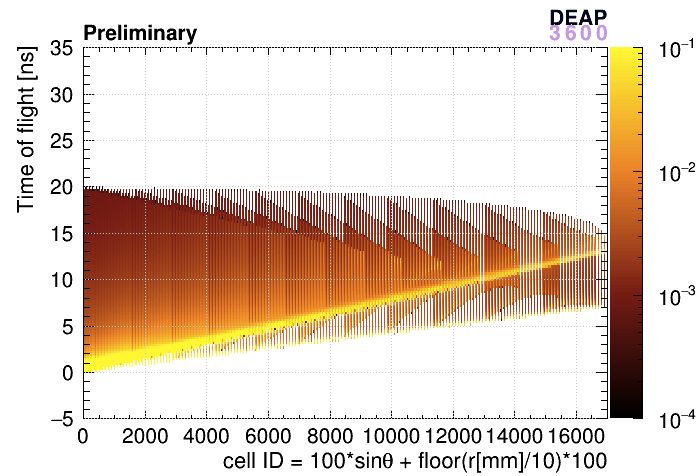}
\subfloat{}\includegraphics[width=0.49\textwidth]{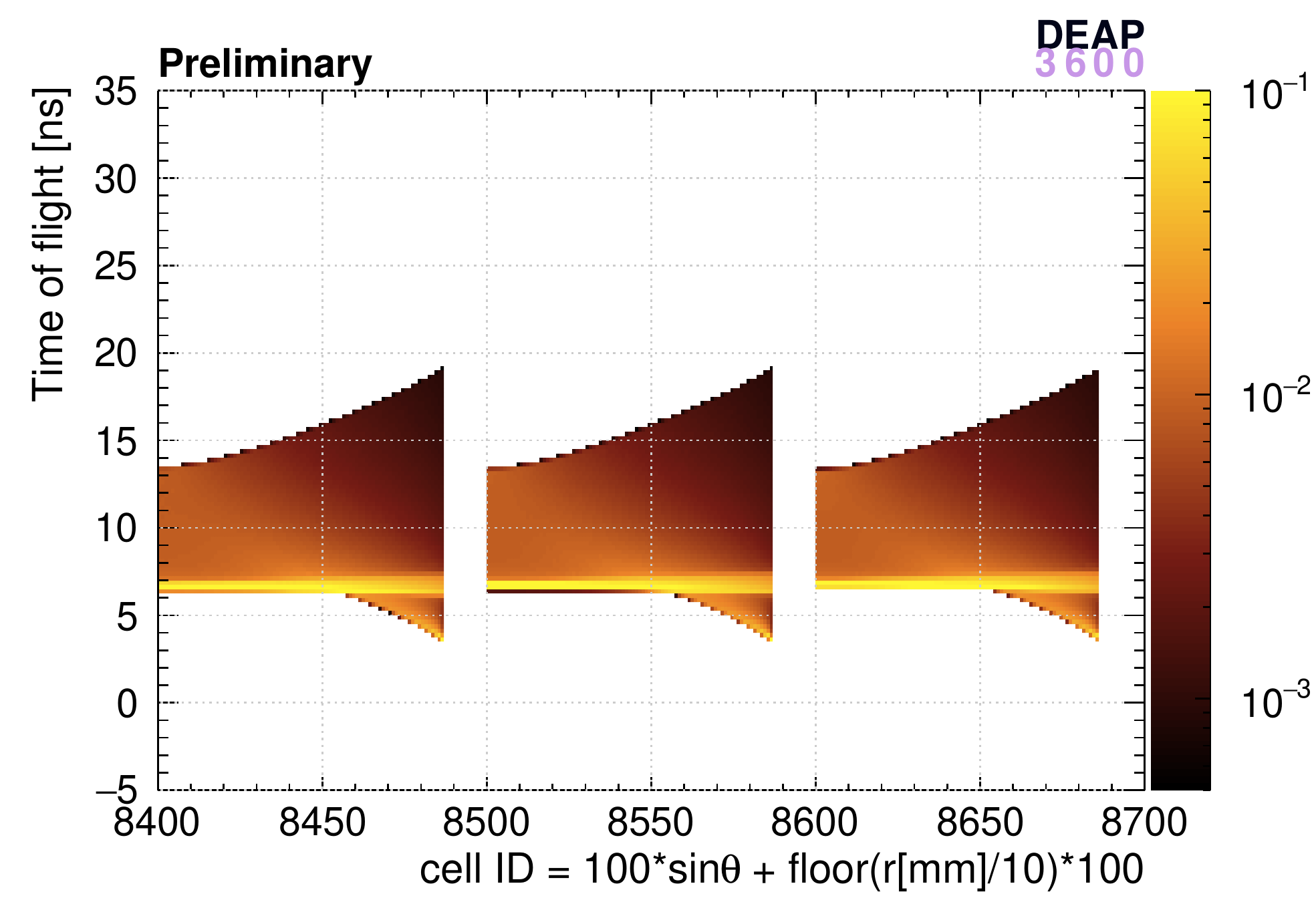}
\caption{PDFs shown as a two-dimensional histogram with respect to time
  of flight and cell ID, with the color scale proportional to probability
  density. $r$ and $\theta$ in the definition of cell ID
  are referred in the right panel of Figure~\ref{fig:scint}. \textit{Left}:
  The histogram including all cells. \textit{Right}:
  The histogram zoomed in to show cells 8,400 to 8,700, representing
  three distance values $r=\SI{840}{\mm}$, $\SI{850}{\mm}$, and
  $\SI{860}{\mm}$ with $\sin\theta$ varying from 0.00 to 0.99 for each
  distance value. }
\label{fig:pdf}
\end{figure}

The probability density functions (PDFs) are constructed for an arbitrary
PMT/LG assembly, in a coordinate system where the origin is the center
of LG front surface. The event position is represented as its distance to
the origin, $r$, and sine of the opposite angle of the perpendicular
distance to the axis, $\sin\theta$, as shown in the right panel
of Figure~\ref{fig:scint}. Then cells (testing points) are deployed at every
\SI{10}{\mm} in $r$, and at every 0.01 in $\sin\theta$. For each cell,
the TOF distribution is determined with numerical calculation, 
integrating Eq.~\ref{eq:tof} with respect to $\vec{x'}$ over the
whole AV surface. The resulting PDFs are shown in
Figure~\ref{fig:pdf} as a two-dimensional histogram, with $y$-axis as
the TOF and $x$-axis as cell ID, with the definition shown in the
label of $x$-axis. In the overall histogram (left panel) the bright strip
represents the time when the majority of the UV light arrives, as it is proportional to
distance. Other features represent the distribution of visible light
emitted by TPB on AV surface. The right panel shows part of this
histogram from cell ID 8,400 to 8,700. The three pieces
represent three distance values $r=\SI{840}{\mm}$, $\SI{850}{\mm}$,
and $\SI{860}{\mm}$ that are typical and equivalent from the center to
the surface of the AV. The spread of each piece represents
$\sin\theta$ varying from 0.00 to 0.99. For large values of
$\sin\theta$, there is light arriving earlier than the majority of the
UV light, because these cells are close to the surface, producing early
TPB light traveling faster than the UV light. The PDF can be noted as
$P(\mathrm{TOF}; r, \sin\theta)$, a function of TOF with cell position
$(r,\sin\theta)$ as parameters. 

The likelihood $\mathcal{L}(t_0,\vec{x_0})$ of a given event time $t_0$
and test position $\vec{x_0}$ is computed as
\begin{equation}
  \label{eq:likelihood}
 \ln\mathcal{L}(t_0,\vec{x_0}) = \sum_{i=1}^{N_{\mathrm{PE}}}\ln
 P(t_i-t_0;\vec{x_0},\mathrm{PMT}_i), 
\end{equation}
where $t_i$ is the time at which the $i$th PE was detected
in channel PMT$_i$; the $N_{\mathrm{PE}}$ in the first 40 ns are
accounted for this calculation. The configuration of $\vec{x_0}$ and
PMT$_i$ is transformed to cell position $(r,\sin\theta)$. This algorithm
returns the values of $\vec{x_0}$ and $t_0$ that maximize
$\mathcal{L}(t_0,\vec{x_0})$.

\section{Validation}
\label{sec:validation}

\begin{figure}[!tbp]
\centering
\includegraphics[width=0.49\textwidth]{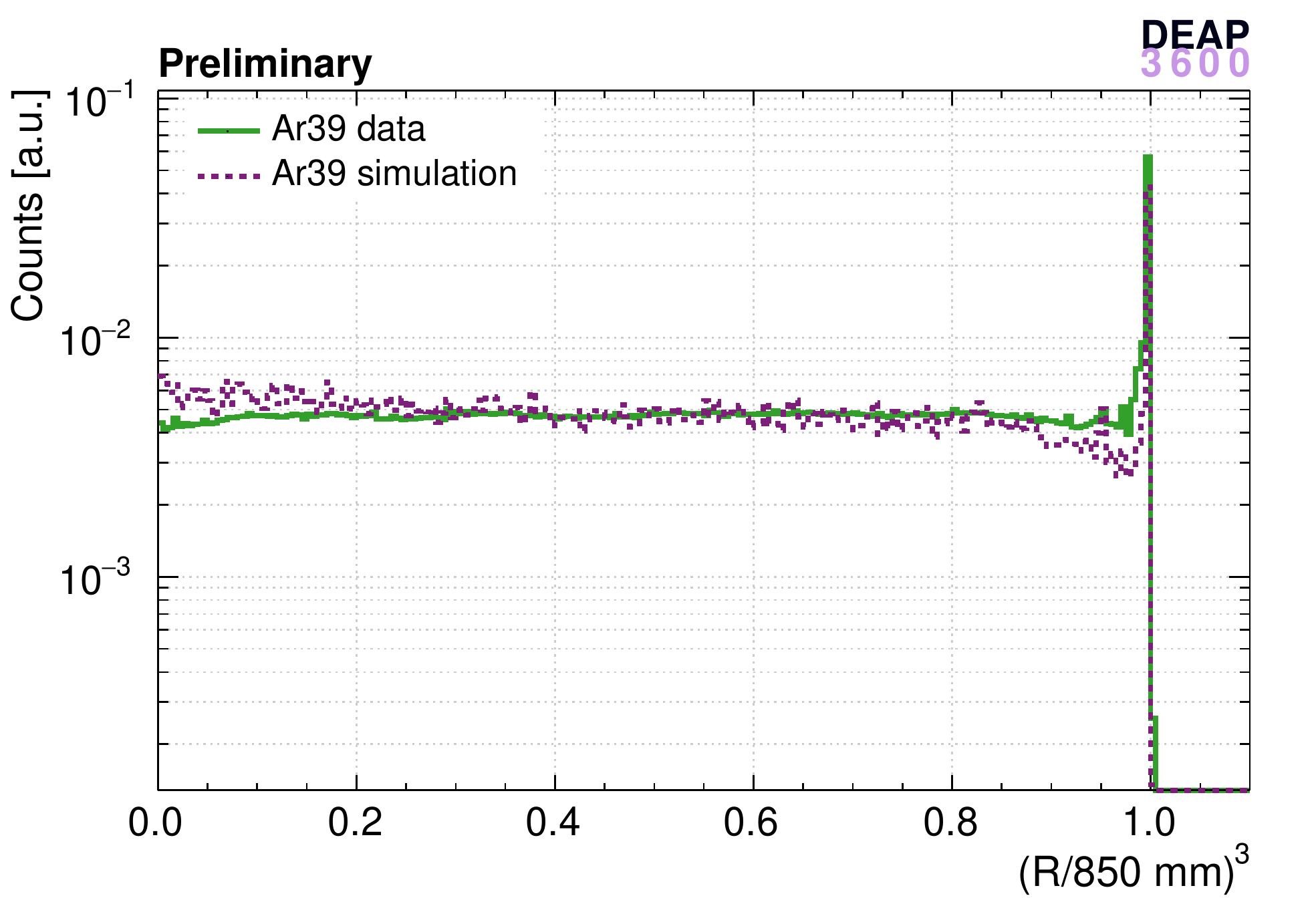}
\includegraphics[width=0.49\textwidth]{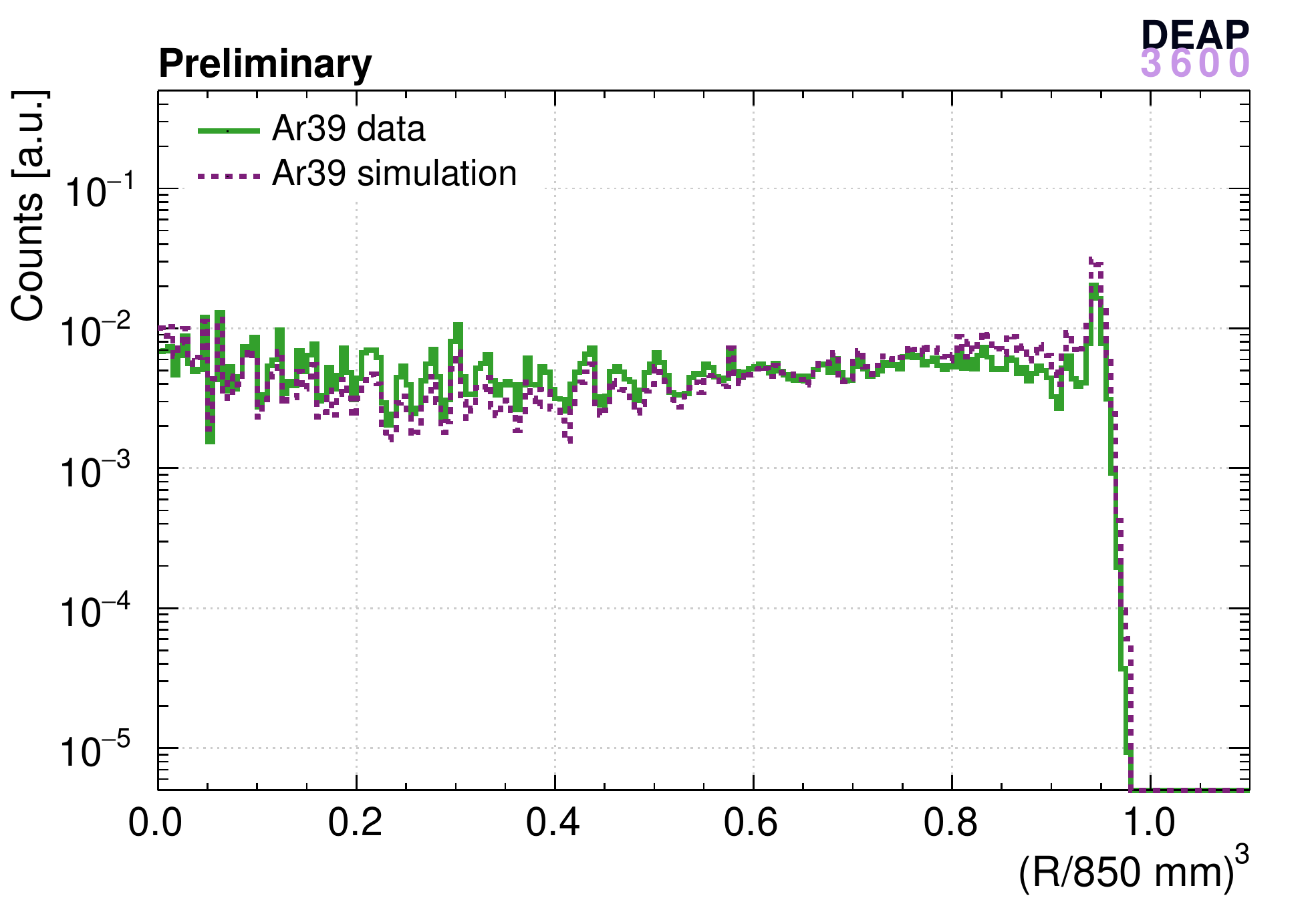}
\caption{
Reconstructed position of $^{39}$Ar $\beta$-decays for data (solid) and Monte
Carlo simulation (dashed) shown in radius cubed, $(R/\SI{850}{\mm})^3$,
 with PE-based algorithm (left) and TOF-based algorithm (right).}
\label{fig:radius}
\end{figure}
The position resolutions for both algorithms are measured with a
data-driven method described in~\cite{willis}. Within the WIMP-search
PE region, near the \SI{630}{\mm} radial cut for fiducialization, the
position resolution for the TOF-based algorithm is typically
40-\SI{50}{\mm}. 

The WIMP-search analysis for the DEAP-3600 experiment relies primarily
on the PE-based algorithm for fiducialization, though it also requires
convergence and agreement between both algorithms. We reconstructed
$^{39}$Ar electronic recoil (ER) events with both algorithms and
showed that the two distributions are uniform in volume. 
Figure~\ref{fig:radius} shows the radius cubed distribution for the
PE-based algorithm on left panel and TOF-based algorithm on
right panel for $^{39}$Ar events. The PE-based algorithm
results in a more uniform distribution in the bulk region, but strongly
biased near the surface, while the distribution with TOF-based
algorithm is more rippled in the bulk but less biased near the
surface. They are both compared with Monte Carlo simulated $^{39}$Ar
$\beta$-decays and show good agreement. The enhancement at the surface
arises because the fitting function is discontinuous in both
algorithms. The discrepancy at zero for the PE-based algorithm is due
to artifacts and is being investigated. 
\begin{figure}[!tbp]
\centering
\includegraphics[width=0.36\textwidth]{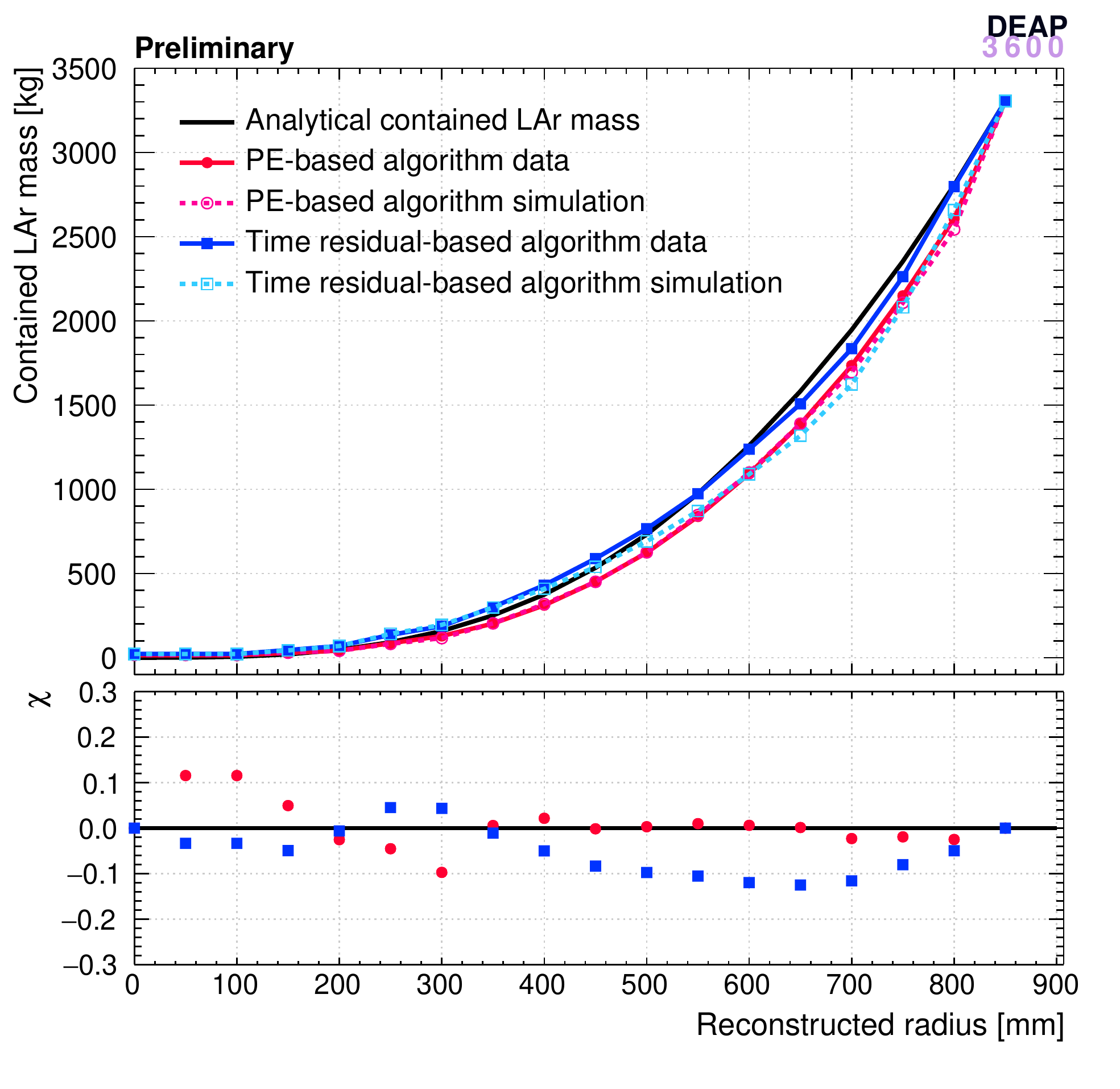}
\includegraphics[width=0.54\textwidth]{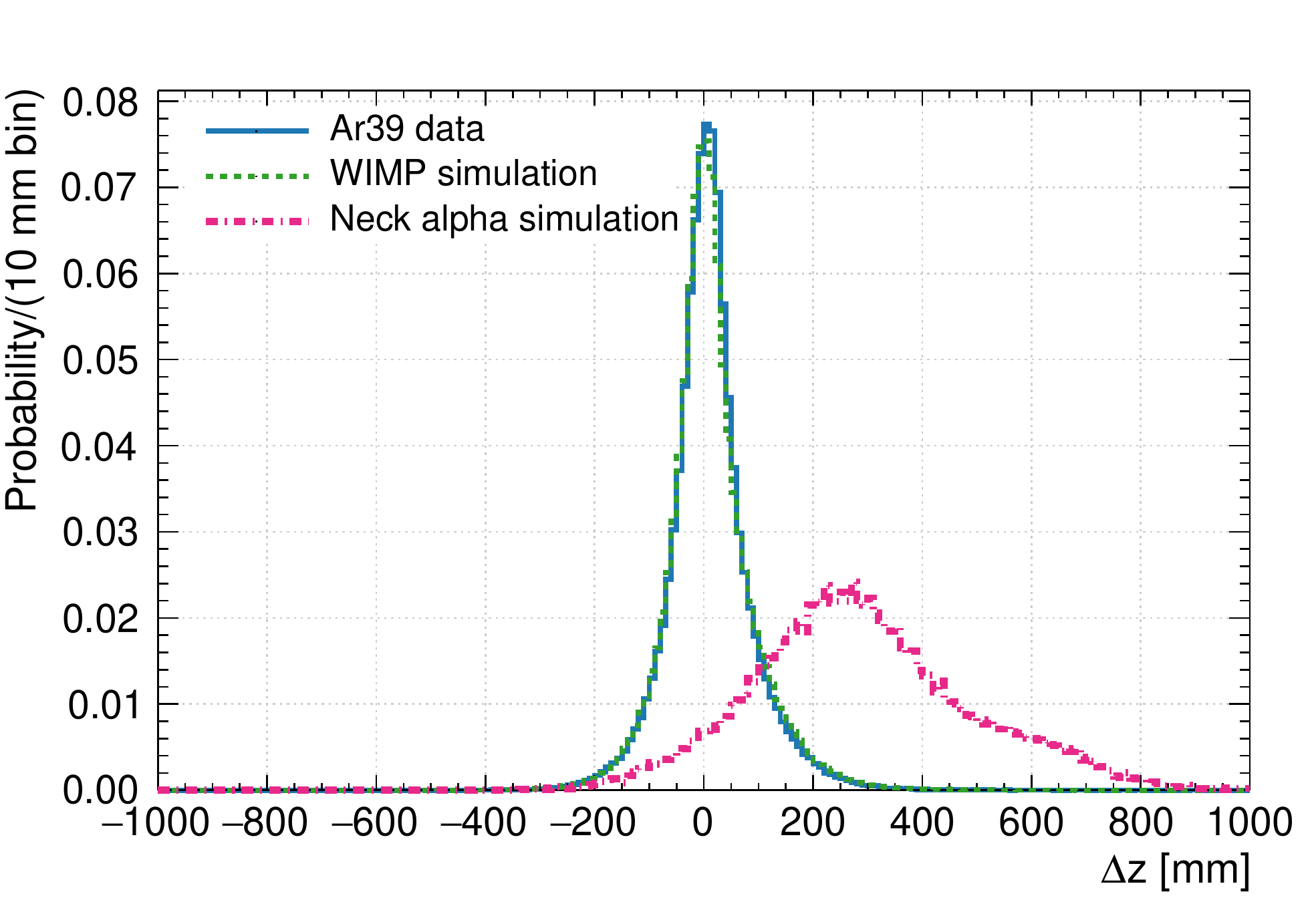}
\caption{\textit{Left:} Estimates from the PE-based and TOF-based
  algorithms of the contained mass of LAr within a radius of
  the reconstructed position for both data and simulation. The bottom
  inset shows the relative difference of simulation and data for each
  algorithm. \textit{Right:} The difference of the vertical positions
  of the TOF-based and PE-based algorithms for $^{39}$Ar data, WIMP
  simulation, and neck alpha simulation. } 
\label{fig:fidmass}
\end{figure}

The left panel of Figure~\ref{fig:fidmass} shows the calculation of
the contained mass of LAr within a
radius of the reconstructed position estimated based on the total
fraction of $^{39}$Ar ER events. Good agreement is achieved between
the mass estimated using both reconstruction algorithms and the
analytical contained LAr mass. In the WIMP search analysis, the
fiducial mass is determined based on the fraction of $^{39}$Ar ER
events in the 95-200 PE range surviving fiducial cuts. 

Both algorithms use the hypothesis that an event is from
a single flash of isotropic light in bulk LAr. When the hypothesis is
true such as for WIMPs or $^{39}$Ar ER events, the two algorithms
should agree. For neck events, however, the light is emitted above the
detector and shadowed by the flowguides, resulting in very different
pulse hitting distribution and TOF distribution. We expect that they
should reconstruct differently. Right panel of
Figure~\ref{fig:fidmass}, published in~\cite{deap2nd}, shows the
difference of the vertical positions of the TOF-based and PE-based
algorithms. It is shown that the reconstructed vertical positions agree within
\SI{35}{\mm} for 50\% of $^{39}$Ar events in data and simulated
WIMPs. For simulated neck $\alpha$-decays, the reconstructed vertical
positions of the two algorithms are very
different, with the TOF-based algorithm systematically reconstructing
these events much closer to the top of the detector than the PE-based
algorithm doing. Including a cut based on the consistency of both
position reconstruction algorithms in our suite of cuts, we achieved
an overall expectation of $0.49^{+0.27}_{-0.26}$ events in ROI
for backgrounds induced by neck $\alpha$-decays in the dataset of 231
live-day WIMP search analysis~\cite{deap2nd}. 

\section{Summary}
\label{sec:con}
DEAP-3600 uses both PE-based and TOF-based algorithms for
position reconstruction. The TOF-based algorithm is
presented in detail. Although it is not used for fiducialization, the
TOF-based algorithm helps remove backgrounds induced by
$\alpha$-decays from the AV neck by comparing the reconstructed
positions with the PE-based algorithm. The TOF-based
position reconstruction uses both spatial distribution and timing
information of the PMT hits. Including the results of the TOF-based
algorithm in our suite of cuts allows us to identify
and remove almost all neck $\alpha$ background events for the dark
matter search.






\end{document}